\documentclass[11pt]{article}

\usepackage{geometry}
\usepackage[utf8]{inputenc}
\usepackage[T1]{fontenc}
\usepackage{amsmath,amssymb,amsfonts}	
\usepackage{graphicx}
\usepackage{color}
\usepackage{xcolor}
\usepackage{xspace}
\usepackage{booktabs}
\usepackage{slashed}

\usepackage[colorlinks=true,pdfstartview=FitV, linkcolor=blue, citecolor=blue,urlcolor=blue]{hyperref} 
\usepackage[sort&compress,numbers,colon,merge]{natbib}

\newcommand{\s}{\ensuremath{\varphi}\xspace}
\makeatletter
\g@addto@macro\bfseries{\boldmath}
\makeatother
\begin{document}
\begin{titlepage}
\vspace*{-1cm}
\phantom{hep-ph/***}
\flushright
\hfil{CPPC-2023-02}

\vskip 1.5cm
\begin{center}
\mathversion{bold}
{\LARGE\bf
  Complementarity of $B\to K^{(*)} \mu \bar \mu$ and $B\to K^{(*)} + \mathrm{inv}$\\[1ex] for searches of GeV-scale Higgs-like scalars 
}\\[3mm]
\mathversion{normal}
\vskip .3cm
\end{center}
\vskip 0.5  cm
\begin{center}
{\large Maksym Ovchynnikov}$^{1}$,
{\large Michael A.~Schmidt}$^{2}$,
{\large Thomas Schwetz}$^{1}$
\\
\vskip .7cm
{\footnotesize
$^{1}$ Institut f\"ur Astroteilchen Physik, Karlsruher Institut f\"ur Technologie (KIT), Hermann-von-Helmholtz-Platz 1, 76344 Eggenstein-Leopoldshafen, Germany\\[0.3cm]
$^{2}$ Sydney Consortium for Particle Physics and Cosmology, School of Physics, The University of New South Wales, Sydney, NSW 2052, Australia\\[0.3cm]

\vskip .5cm
\begin{minipage}[l]{.9\textwidth}
\begin{center}
\textit{E-mail:}
\tt{maksym.ovchynnikov@kit.edu}, \tt{m.schmidt@unsw.edu.au}, \tt{schwetz@kit.edu}
\end{center}
\end{minipage}
}
\end{center}
\vskip 1cm
\begin{abstract}
The rare decays $B^+\to K^+ \mu\bar \mu$ and $B^0\to K^{*0} \mu\bar\mu$ provide the strongest constraints on the mixing of a light scalar with the Higgs boson for GeV-scale masses. The constraints sensitively depend on the branching ratio to muons. Additional decay channels like an invisible partial width may substantially weaken the constraints. This scenario will be probed at Belle II in $B\to K^{(*)} + \mathrm{inv}$. We illustrate the complementarity of scalar decays to muons and invisible decays using the currently available results of LHCb and BaBar. We provide two simple model realisations providing a sizeable invisible scalar width, one based on a real scalar and one based on a $U(1)_{B-L}$ gauge symmetry. In both examples the scalar decays into heavy neutral leptons which can be motivated by the seesaw mechanism for neutrino masses.
\end{abstract}
\end{titlepage}

\section{Introduction}
\label{sec:intro}

Light GeV-scale Higgs-like scalars \s occur in several well-motivated extensions of the Standard Model (SM), in which a SM gauge group singlet scalar field $\varphi$ couples to the Higgs via a cubic or quartic interaction. Examples include a mediator between a dark sector and the SM~\cite{Pospelov:2007mp,Krnjaic:2015mbs}, a light inflaton~\cite{Shaposhnikov:2006xi,Bezrukov:2009yw,Bezrukov:2013fca}, the pseudo-Goldstone bosons associated with the breaking of scale-invariance in a classically scale-invariant model~\cite{Foot:2011et}, or light scalars to explain different low-energy anomalies, see e.g.~\cite{Egana-Ugrinovic:2019wzj,Dev:2019hho,Abdullahi:2020nyr}. The light scalar could be the scalar breaking $B-L$ symmetry~\cite{Davidson:1978pm,Marshak:1979fm}, which has been studied in e.g.~\cite{Dev:2021qjj}. 

The phenomenology of GeV-scale Higgs-like scalars has been recently studied in~\cite{Batell:2009jf,Clarke:2013aya,Boiarska:2019jym} (see also the review \cite{Ferber:2023iso} and \cite{Balaji:2022noj} for recent stellar limits). Below the scale of the electroweak symmetry breaking, the interaction of the scalars with the SM particles may be generically described by the two independent interactions: the mixing with the Higgs boson $h$ (parameterised by the mixing angle $\theta$), and the trilinear coupling $h \s\s$ which gives rise to an invisible Higgs decay. The mixing coupling makes the scalar unstable: it may decay into a pair of leptons or into hadrons.

The most stringent constraints on Higgs-like scalars with masses $0.3\,\mathrm{GeV}\lesssim m_\varphi<m_B-m_K$ (see e.g.~\cite{Beacham:2019nyx})
come from the LHCb displaced vertex search for $B\to K \varphi(\to \mu\bar\mu)$~\cite{LHCb:2015nkv,LHCb:2016awg} which constrains the scalar mixing down to $\theta \simeq 10^{-4}$ depending on the scalar mass.
The corresponding search for long-lived particles decaying to a pair of light leptons, $e\bar e, \mu\bar\mu$ or light mesons $\pi\pi, KK$~\cite{Dreyer:2021aqd,Belle-II:2023ueh} are currently less sensitive. 
However, these constraints are subject to the assumption that the scalar dominantly decays visibly into SM particles which may not hold in general (if additional couplings apart from the mixing exist). Thus it is crucial to also consider searches for invisible decays of the scalar. 

\begin{table}[b!]\centering
  \begin{tabular}{lcccc}\toprule
    decay & upper bound & SM & \multicolumn{2}{c}{Belle II sensitivity~\cite{Belle-II:2022cgf}} \\
          && &$1 \mathrm{ab}^{-1}$ 
           & $50\mathrm{ab}^{-1}$ \\\midrule
          $\mathrm{Br}(B^+\to K^+ +\mathrm{inv})$ &$<1.6\times 10^{-5}$~\cite{BaBar:2013npw} & $4.6\times 10^{-6}$ & 0.55(0.37)
                                             & 0.11 (0.08)\\
                                            $\mathrm{Br}(B^0\to K_S^{0} +\mathrm{inv})$ &$<2.6\times 10^{-5}$~\cite{Belle:2017oht} & $2.1\times 10^{-6}$ & 2.06 (1.37) 
                                                 & 0.59(0.40) \\
                                                $\mathrm{Br}(B^+ \to K^{*+} + \mathrm{inv})$ & $ < 4.0\times 10^{-5}$~\cite{Belle:2013tnz}& $11\times 10^{-6}$ & 2.04(1.45)
                                                  & 0.53(0.38)\\
                                                 $\mathrm{Br}(B^0\to K^{*0} +\mathrm{inv})$ & $<1.8\times 10^{-5}$~\cite{Belle:2017oht} & $10\times 10^{-6}$ & 1.08(0.72) 
                                                & 0.34(0.23) \\
   \bottomrule
 \end{tabular}
 \caption{Upper bounds on the branching ratios $B\to K+\mathrm{inv}$ and the baseline (improved) expected sensitivities of Belle II to the signal strength relative to the SM assuming the kinematic distribution of a 3-body decay $B\to K\nu\bar\nu$. The SM prediction has been obtained using the expressions in~\cite{Felkl:2021uxi} with the form factors~\cite{Gubernari:2023puw} using the Bharucha-Straub-Zwicky (BSZ) parametrization~\cite{Bharucha:2015bzk} and the recent lattice result for $B\to K$~\cite{Parrott:2022rgu}. In addition, we add a 20\% correction for $B\to K^* \nu\bar\nu$ due to finite width effects of $K^*$~\cite{Descotes-Genon:2019bud}.
 They agree within errors with \cite{Becirevic:2023aov}, which predicts a slightly enhanced branching ratio for $B^+\to K^+\nu\bar\nu$ and a reduced one for $B^0\to K^{*0}\nu\bar\nu$. Compared to~\cite{Straub:2018kue,david_straub_2023_7859312,Altmannshofer:2009ma,Brod:2010hi,Buras:2014fpa}, the branching ratios for $B\to K \nu\bar\nu$ are 5 \% larger and the ones for $B\to K^* \nu\bar\nu$ 10\% smaller due to the difference in form factors.
 }
\label{tab:BKnunu}
\end{table}

The $B$ factories BaBar, Belle and Belle II searched for $B\to K^{(*)} +\mathrm{inv}$ and reported upper limits on the different decay channels which are listed in Tab.~\ref{tab:BKnunu}. 
Belle II is expected to measure all four decay channels including the polarisation fraction $F_L$ of the $K^*$~\cite{Belle-II:2018jsg}, which may be used to search for additional invisible final states like heavy neutral leptons (HNLs)~\cite{Kim:1999waa,Felkl:2021uxi} or light invisibly-decaying scalars~\cite{Bird:2004ts,Altmannshofer:2009ma, He:2022ljo}, which may constitute dark matter. 
In fact, the Belle II collaboration already published their first analysis~\cite{Belle-II:2021rof}. 
A simple weighted average indicates the branching ratio $\mathrm{Br}(B^+\to K^+ +\mathrm{inv}) = (1.1\pm 0.4)\times 10^{-5}$, which is $1.4\sigma$ in excess over the SM prediction.

The complementarity of displaced and invisible searches has also recently been highlighted for Belle II.
Ref.~\cite{Filimonova:2019tuy} studied a Higgs-like scalar decaying to dark sector particles. It stressed the benefit of searching for displaced pairs of leptons or light mesons in the Belle II detector. They conclude that Belle II is able to probe mixing angles down to $10^{-5}$. 
Ref.~\cite{Ferber:2022rsf} studied axion-like particles. While displaced vertex searches are currently more sensitive for masses above the muon threshold, invisible decay searches are more sensitive for lighter masses. 

In this paper, we study the bounds on the light scalar mixing with the SM Higgs under the assumption of a sizeable invisible width of the scalar $\Gamma_{\rm inv}$. We show how the limits from $B\to K\mu\bar\mu$ become weaker if the scalar invisible width is increased. At the same time bounds from $B\to K +\mathrm{inv}$ become relevant and for sufficiently large values of $\Gamma_{\rm inv}$ they will dominate the bound on the scalar/Higgs mixing. We illustrate this complementarity by using current LHCb bounds on $B\to K\mu\bar\mu$~\cite{LHCb:2015nkv,LHCb:2016awg} and the BaBar bound on $\mathrm{Br}(B^+\to K^+ +\mathrm{inv})$~\cite{BaBar:2013npw}. We will give also two simple examples for light new physics, where the required values for $\Gamma_{\rm inv}$ can be achieved by letting the scalar decay into HNLs. 

The paper is structured as follows. In Sec.~\ref{sec:model} we introduce the couplings of the scalar and the relevant decay modes. In Sec.~\ref{sec:overview} we provide an overview of searches for scalars at $B$ factories. The complementarity of $B\to K \mu\bar\mu$ and $B\to K +\mathrm{inv}$ is presented as the main result in Sec.~\ref{sec:results}. In Sec.~\ref{sec:examples} we provide a few examples for models with a sizeable invisible decay width, before concluding in Sec.~\ref{sec:conclusions}. Technical details are collected in the appendix.

\section{Phenomenology of GeV-scale Higgs-like scalars in a nutshell}
\label{sec:model}
We consider a light real scalar field\footnote{It may be the real component of a complex scalar field like in the gauged $B-L$ extension of the SM with GeV-scale mass $m_\varphi$.} $\varphi$, which mixes with the Higgs with mixing angle $\theta$. In the minimal scenario, $\theta$ and the scalar mass $m_{\varphi}$ control every observable such as the scalar lifetime $\tau_{S}\propto f(m_{\varphi})\theta^{-2}$ and the partial branching ratios.

Because of the mixing, the structure of the scalar interaction with SM particles is similar to those for the Higgs, but the mixing angle suppresses the couplings. This way, $\varphi$ couples to all SM fermions at tree level proportional to their respective masses as $(m_f \sin\theta /v)$. These tree-level interactions generate effective couplings to other SM particles such as gluons, photons, and the bound states such as nucleons~\cite{Boiarska:2019jym}. They differ from those of the Higgs boson not only by $\theta$ but also by mass which determines the scale associated with the couplings.

Thus, the scalar decays to kinematically-accessible SM lepton pairs with partial decay width\footnote{We assume small scalar mixing angle $\sin\theta\ll 1$ and thus use $\cos\theta\approx 1$.}  
\begin{equation}
\Gamma(\varphi\to f\bar f)= \frac{\sqrt{2}G_F m_f^2 m_\varphi \sin^2\theta}{8\pi}  \left(1-\frac{4m_f^2}{m_\varphi^2}\right)^{3/2} \;.
\label{eq:decay-width-leptons}
\end{equation}
Decays into hadrons are more complicated. Their description depends on the scalar mass which sets a characteristic energy scale of the process. For $m_{\s}\gg \Lambda_{\text{QCD}} = \mathcal{O}(1\text{ GeV})$, they may be described inclusively by the decay of the scalar into a quark-antiquark pair; the corresponding decay width is given by Eq.~\eqref{eq:decay-width-leptons} with the additional colour factor $N_c=3$. 
For lower masses, $m_{\s}\simeq \Lambda_{\text{QCD}}$, the perturbative QCD description breaks down, and hadronic decays must be treated exclusively, i.e., into different mesons.  

The lightest possible hadronic decay is into a pair of pions, $\varphi \to \pi \pi$. It may be described using chiral perturbation theory, which, however, quickly becomes unreliable just above the decay threshold due to strong interactions of the pions~\cite{Donoghue:1990xh}. Alternatively, the calculation of the decay form-factor may be performed using the method of dispersion relations, see~\cite{Winkler:2018qyg} for an overview. It was realised (see, e.g.,~\cite{Monin:2018lee}) that the approach suffers from theoretical uncertainties that may significantly change the results. Recently,~\cite{Gorbunov:2023lga} has calculated the decay width using experimental data for the gravitational pion form-factors with an uncertainty of about a factor $\mathcal{O}(2)$.

\begin{figure}
    \centering
    \includegraphics[width=0.5\textwidth]{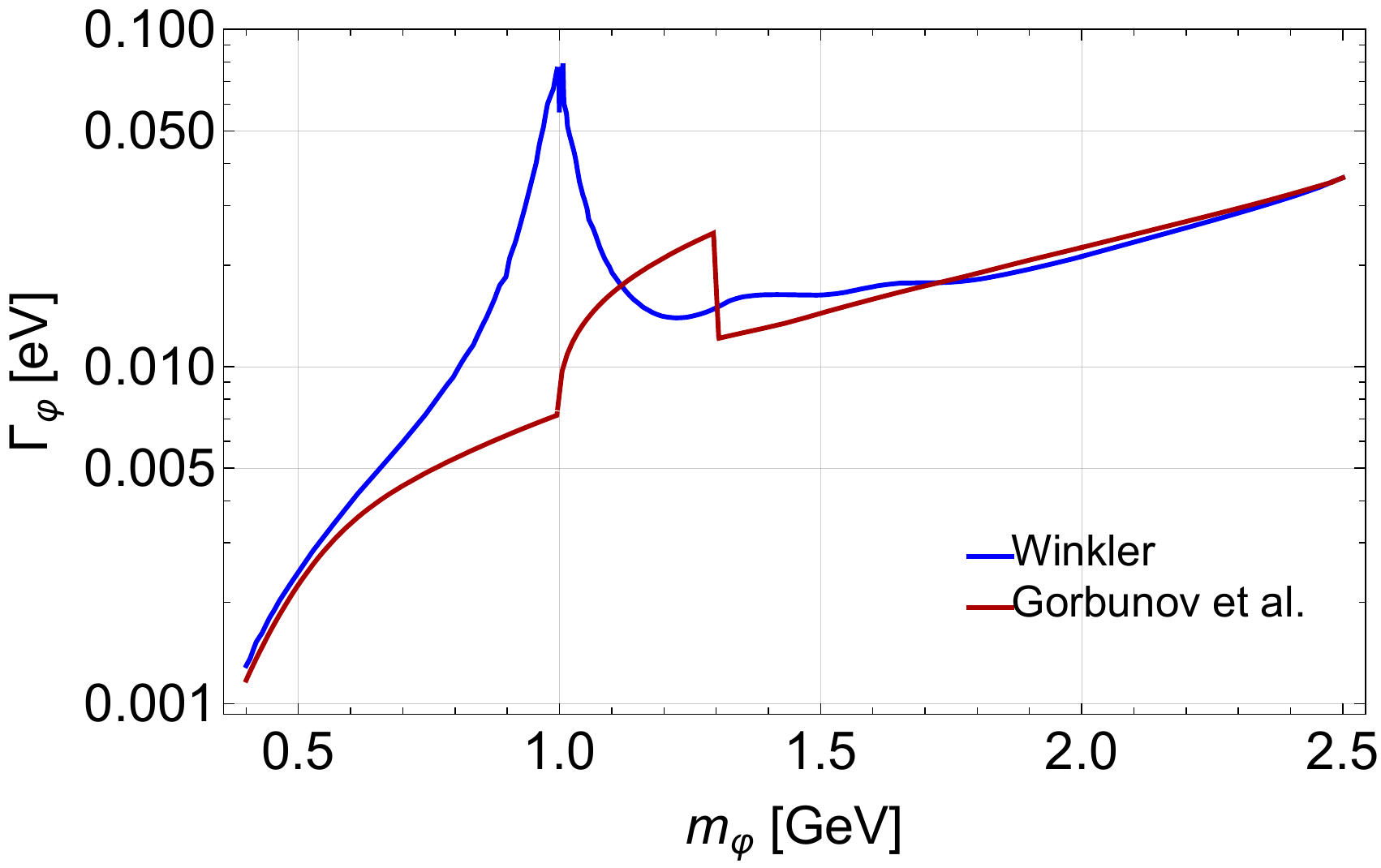}~\includegraphics[width=0.5\textwidth]{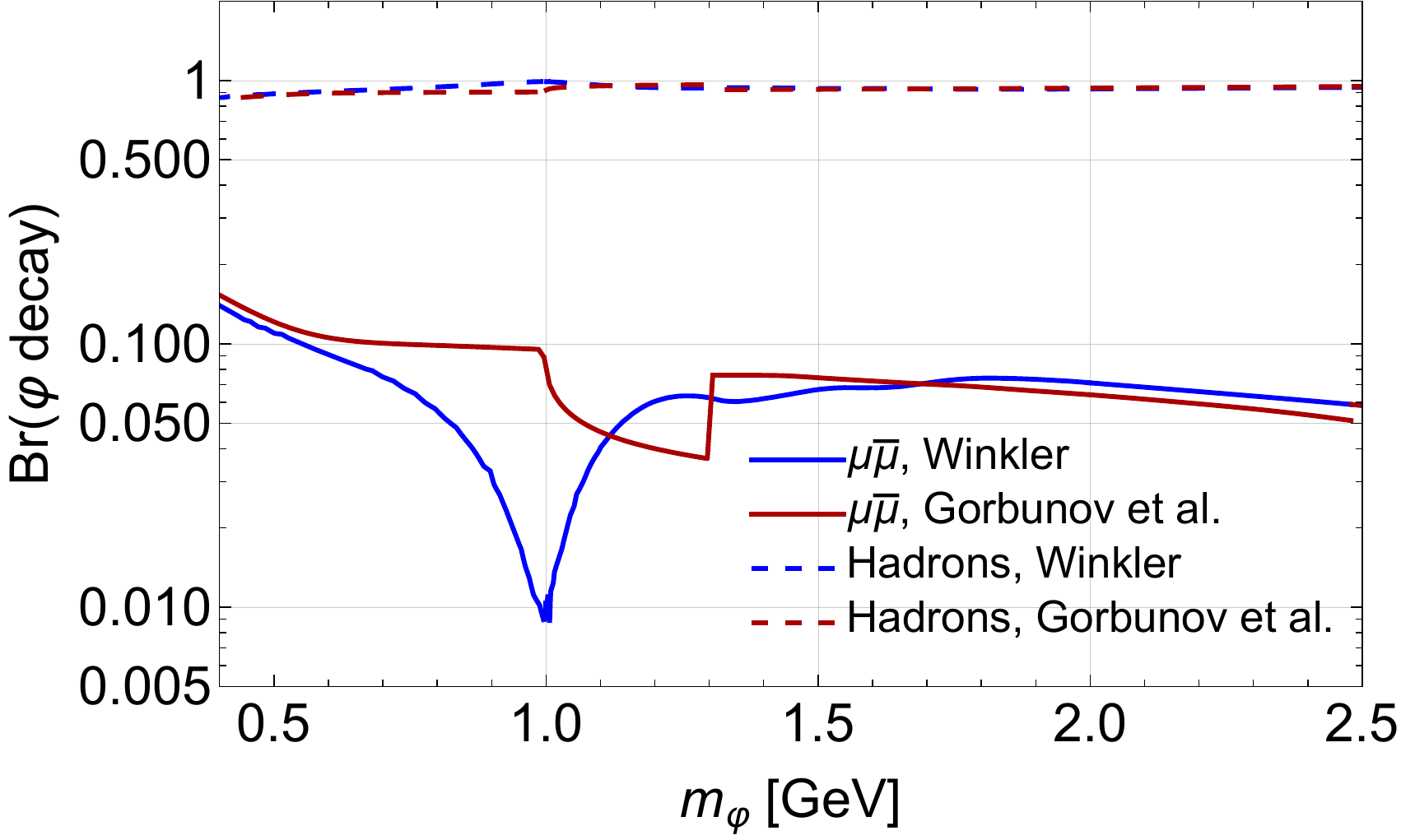}
    \caption{Decays of a GeV-scale scalar \s in the minimal model described by the two parameters -- the scalar mass $m_{\s}$ and the mixing angle with the SM Higgs boson $\theta$. The left panel shows the decay width of the scalar at $\sin^2\theta = 10^{-4}$, while the right panel the branching ratio of its decays into pairs of muons or hadrons. The blue lines indicate the result obtained in~\cite{Winkler:2018qyg}, with included NLO corrections for decays into quarks and gluons. The red lines show the calculations of~\cite{Gorbunov:2023lga}. The difference between them indicates the uncertainty in the description of hadronic decays of the scalar (see text for details).
    }
    \label{fig:scalar-lifetime}
\end{figure}

Other hadronic decays of \s, e.g., into a pair of kaons and multihadronic states, also suffer from similar problems. In particular, there is no clear way to describe the transition between the exclusive and inclusive approaches. Refs.~\cite{Winkler:2018qyg,Gorbunov:2023lga} match the two regimes at different scalar masses -- $2\text{ GeV}$ and $1.3\text{ GeV}$, respectively.  Motivated by this issue, in Fig.~\ref{fig:scalar-lifetime} we show the lifetime and partial branching ratios of \s following the descriptions from~\cite{Winkler:2018qyg} and~\cite{Gorbunov:2023lga}, interpreting the difference between them as the uncertainty in the decay width.
Depending on the scalar mass, the difference between the widths may be as large as an order of magnitude. We will comment on the impact of this uncertainty on searches for scalars considered in this work in Sec.~\ref{sec:results}.

Let us now discuss the scalar production channels. Depending on the facility, the main channels are~\cite{Boiarska:2019jym} proton bremsstrahlung or decays of various particles: 2- and 3-body decays of kaons and $B$ mesons $B\to \s + \text{other}$, or a 2-body decay of the Higgs boson $h\to \s\s$.
The production channel of main interest in this work is the decay of $B$ mesons into a scalar and a strange meson:
\begin{equation}
    B\to \s+X_{s}, \quad X_{s} = K,K^{*},K_{0}^{*},K_{1}(1270),K_{1}(1400),K_{2}^{*},\dots
\end{equation}
The decay vertex originates from a flavour-changing neutral current coming from weak 1-loop contributions~\cite{Willey:1982mc,Leutwyler:1989xj} (see also Appendix~\ref{app:formfactor} for details).

In the limit of small scalar masses $m_{\s}\ll m_{B}$ and $\sin^{2}\theta\ll 1$, the total exclusive branching ratio is~\cite{Boiarska:2019jym} (see also Appendix~\ref{app:formfactor})
\begin{equation}
    \sum_{X_{s}}\text{Br}(B\to X_{s}+\s) \approx 3.3\, \sin^{2}\theta
\end{equation}
Decays into $K$ consist of only around $10\%$ of this number:
\begin{equation}
    \text{Br}(B\to K+\s)\approx 0.4\,\sin^{2}\theta, \quad m_{\s}\ll m_{B}
\end{equation}
Despite this small branching ratio, this channel is attractive from the experimental point of view. The main reason is that, unlike the heavier resonances $(K^*, K^*_0, K_1,...)$, $K$ is stable on the experiment scales, so the kinematics reconstruction is simple: it only requires to reconstruct the kaon itself. The other resonances are short-lived and decay into a kaon plus other particles such as pions and photons, which require additional signal selection and reconstruction.
Because of this, searches for new physics via the $B\to K\mu\bar\mu$ channel typically give the strongest constraint.

\section{Overview of searches for scalars at \texorpdfstring{$B$}{B} factories} \label{sec:overview}

Two types of searches at $B$ factories -- BaBar, Belle/Belle II, LHCb -- may be applied to dark scalars: $B\to K+\text{visible}$, or $B\to K+\text{invisible}$. 

The first type corresponds to the production $B\to K+\s$, with \s subsequently decaying into visible particles within the detector. In general, such decays may be $\s\to \mu\bar\mu, \s\to \pi\pi, \s \to KK$, as well as decays into jets -- quarks and gluons. The number of events for this signature (without taking into account further selection and reconstruction) is proportional to
\begin{align}
    \mathrm{BR}(B^+\to K^+ \varphi) \,\mathrm{BR}(\varphi\to \text{visible})\, \left[1- P(r_{\rm det}|\beta\gamma)\right]\;,
    \label{eq:visible-scaling}
\end{align}
where $P(r_{\rm det}|\beta\gamma)=\exp[-r_{\rm det}/\beta\gamma c\tau_{\s}]$ denotes the probability to decay outside of the detector of size $r_{\rm det}$\footnote{For LHCb, $r_{\rm det}=0.6$m~\cite{LHCb:2016awg}.}, and $\mathrm{BR}(\varphi\to \text{visible})$ is the branching ratio of decays of \s into visible particles. The scalar \s is produced on-shell and thus the process can be considered as a series of 2-body decays with 
\begin{align}\label{eq:3body-decay}
  \Gamma(B\to K^{(*)}  \varphi (\to f\bar f)) & = \int_0^\infty \frac{dq^2}{2\pi} \frac{\left[\Gamma(B\to K^{(*)}\varphi)2m_\varphi \Gamma(\varphi\to f\bar f) \right]_{m_\varphi^2\to q^2}}{(q^2-m_\varphi^2)^2 + m_\varphi^2\Gamma^2_\varphi}\\\nonumber
                            & \stackrel{\Gamma_\varphi\ll m_\varphi }{\longrightarrow} \Gamma(B\to K^{(*)} \varphi) \mathrm{BR}(\varphi\to f\bar f)
\end{align}
where $q^2$ denotes the 4-momentum squared of the scalar and the narrow width approximation has been employed in the last line, which is a good approximation for the relevant parameter space. As a result, the invariant mass distribution of the visible particles (if they are fully reconstructed) would have a peak at $m_{\rm inv} = m_{\s}$ with a width due to the finite resolution of the 4-momenta reconstruction, which may be used to reduce the background efficiently.

From Fig.~\ref{fig:scalar-lifetime}, we conclude that the most probable decay of a GeV-scale scalar is into hadrons, in particular into a pair of pions, kaons, or jets such as two gluons. The decay into muons is significantly suppressed. However, the latter may give the cleanest decay due to better reconstruction capabilities for muons and lower backgrounds.

For the minimal scalar model with only two parameters $m_{\s}$ and $\theta$, the most stringent constraints on a GeV-scale Higgs-like scalar come from LHCb~\cite{LHCb:2015nkv,LHCb:2016awg}. Namely, Ref.~\cite{LHCb:2016awg} searched for displaced vertices in the decays $B^+\to K^+  \varphi (\to \mu\bar\mu)$~\cite{LHCb:2016awg}, while the work~\cite{LHCb:2015nkv} considered $B^0 \to K^{*0}  \varphi(\to \mu\bar\mu)$. 
Belle II~\cite{Belle-II:2023ueh} also searched for $B^+\to K^+ \varphi(\to \mu\bar\mu)$.
The strongest constraint is placed by~\cite{LHCb:2016awg} which we thus use it in the following analysis.
It placed constraints on the scalar mixing angle as a function of the scalar mass $m_\varphi$ and the lifetime $\tau_\varphi$ down to  $\sin\theta\lesssim 10^{-4}$.

\begin{figure}[t]
    \centering
    \includegraphics[width=0.5\textwidth]{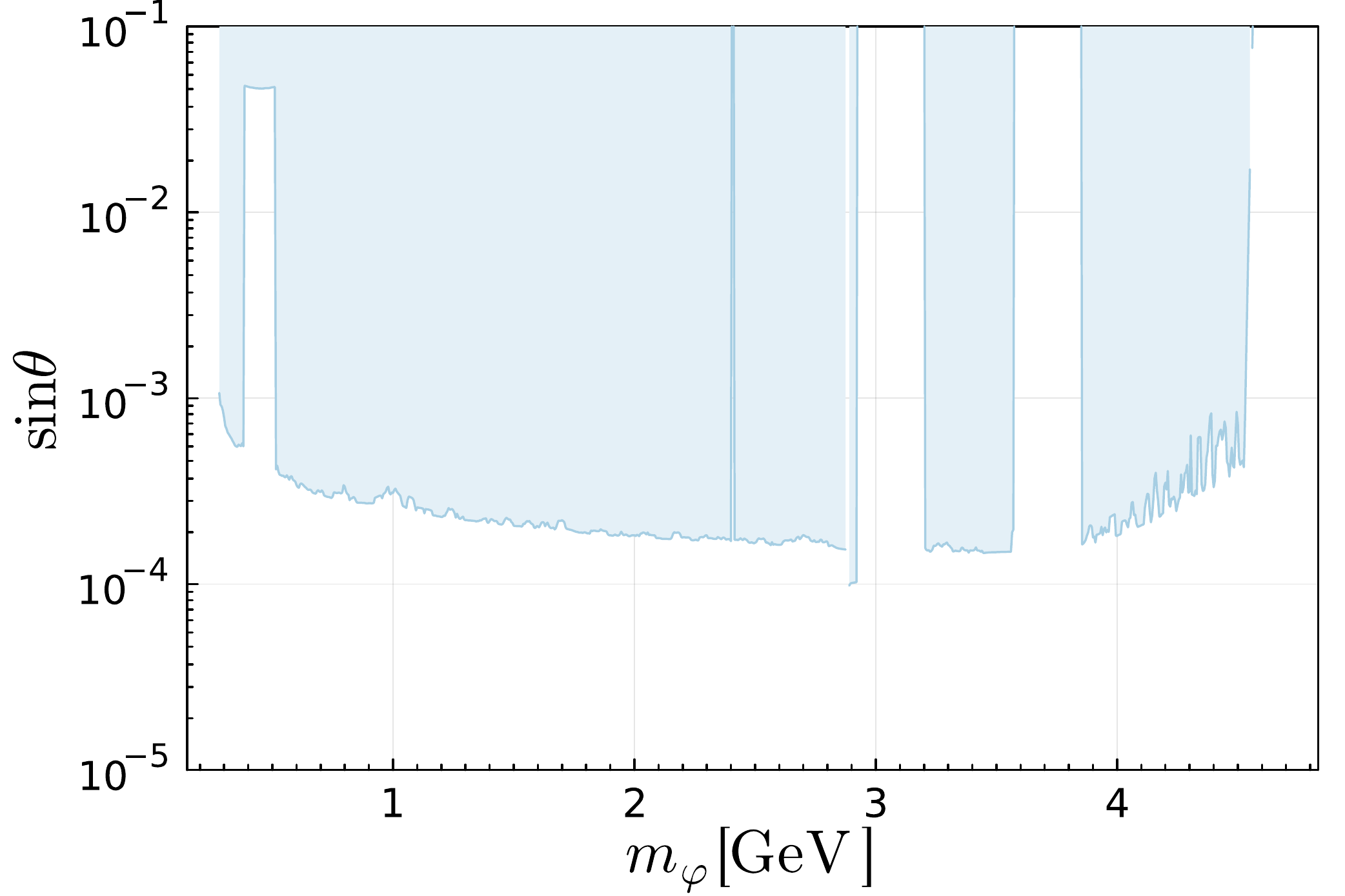}~\includegraphics[width=0.5\textwidth]{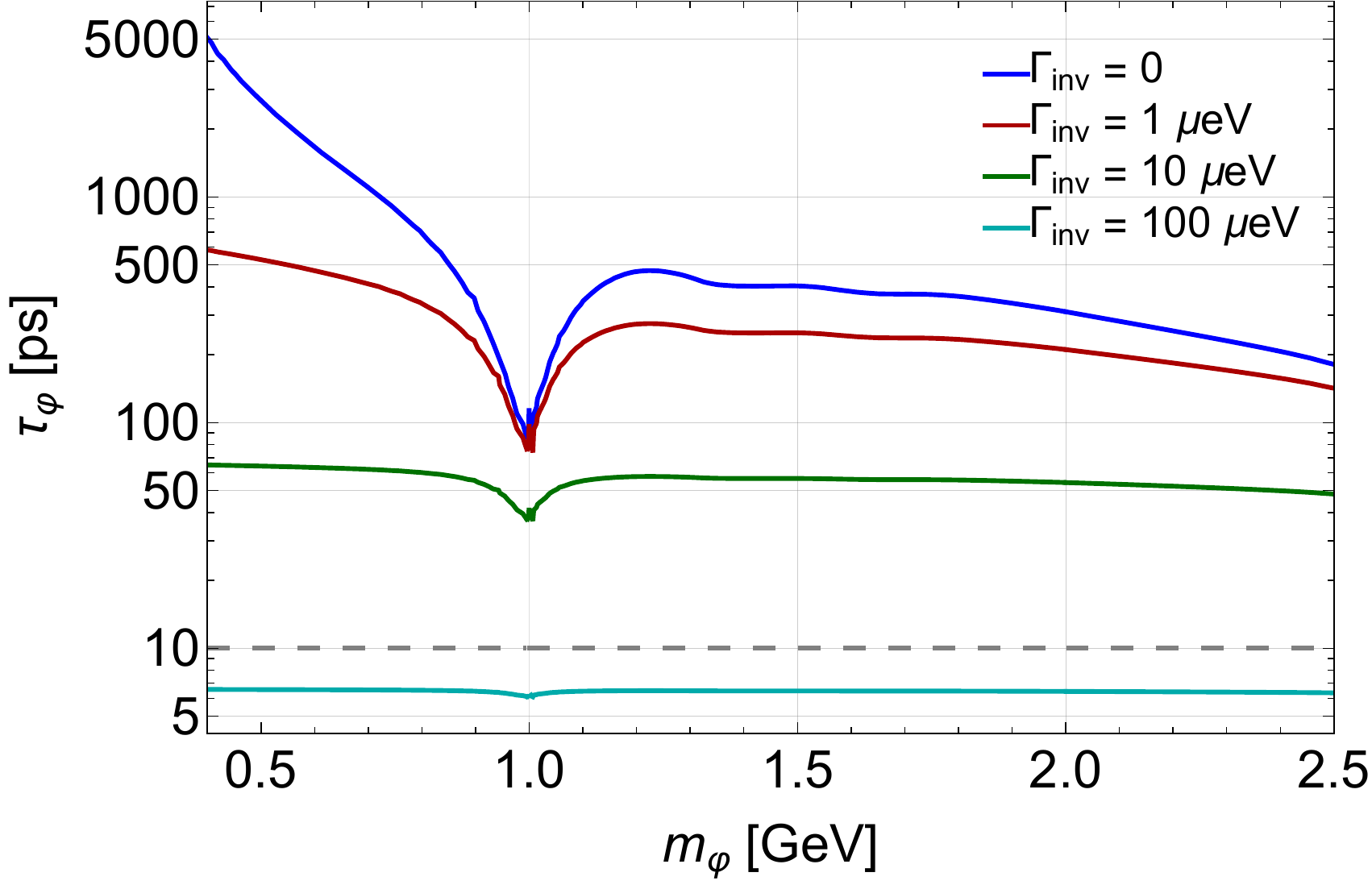}
    \caption{Left panel: constraints on the scalar parameter space coming from searches for $B\to K \mu\bar\mu$ at LHCb for $\Gamma_{\rm inv} = 0$ (see text for details). Right panel: the scalar lifetime at $\sin\theta = 10^{-4}$ as a function of its mass for various values of the decay width $\Gamma_{\rm inv}$, which, being summed with the $\theta$-controlled width gives the total decay width of the scalar. The dashed grey line denotes the lifetime $\tau_{\phi} = 10\text{ ps}$, above which the scalar lifetime is too large to decay inside the detector frequently, such that the decay probability scales as $\tau_{\s}^{-1}$ and the event rate becomes independent of the total decay width.}
    \label{fig:constraints-lhcb}
\end{figure}

We extracted the constraint on BR($B^+\to K^+ \varphi$)BR($\varphi\to\mu\bar\mu$) as a function of the scalar mass $m_\varphi$ and lifetime $\tau_\varphi$ from Fig.~4 in~\cite{LHCb:2016awg} which is model-independent. In Fig.~\ref{fig:constraints-lhcb} (left) we use this constraint to set a limit on the scalar mixing, where we assume the scalar decay width as in~\cite{Winkler:2018qyg} but including NLO corrections to decay widths into quarks and gluons. 
The gaps at several masses are caused by the exclusions in the searched mass range due to the contribution of SM resonances $K^{0}_{S}, J/\psi,\psi(2S)$, and $\psi(3770)$. Interestingly, the lower bound of the excluded region does not depend on the total decay width of the scalar, being determined only by $\Gamma({\s \to \mu\bar\mu})$; in particular, there is no drop of the sensitivity at $m_{S}\simeq 1\text{ GeV}$ observed in Fig.~\ref{fig:scalar-lifetime} for $\text{Br}(\s\to \mu\bar\mu)$. The reason for this is that scalars with $\sin\theta$ close to the lower bound are very long-lived, with the characteristic decay length $c\tau_{\s}\langle\beta\gamma_{\s}\rangle$ exceeding much the geometric size of the detector. As a result, the decay probability in~\eqref{eq:visible-scaling} behaves as $1- P(r_{\rm det}|\beta\gamma) \propto \tau_{\s}^{-1}$, and the product $\text{Br}(\s\to \mu\bar\mu)\tau_{\s}^{-1}$ is just $\Gamma(\s \to \mu\bar\mu)$. Indeed, the scalar's proper lifetime at the vicinity of 1 GeV and $\sin\theta\sim 10^{-4}$ is $\tau_{\s}\gtrsim \mathcal{O}(100)\text{ ps}$, which corresponds to the ``extremely displaced region'' in the LHCb search $\tau \gg 10\text{ ps}$. Because of this, in the minimal scalar model, the lower bound does not depend on the description of the hadronic decays of the scalar (remind Fig.~\ref{fig:scalar-lifetime}). 

The situation may be different if the total scalar decay width has contributions from $\theta$ (given by the decay width $\Gamma^{(\theta)}_{\s}$) as well as another contribution, for instance, an invisible decay width $\Gamma_{\text{inv}}$:
\begin{equation}
\Gamma_{\s,\text{tot}}=\Gamma^{(\theta)}_{\s}  +\Gamma_{\rm inv}
\;.
\end{equation}
If the latter is sufficiently large to be comparable with $\Gamma^{(\theta)}_{\s}$, the lifetime becomes small enough such that all scalars decay inside the detector, see Fig.~\ref{fig:constraints-lhcb} (right).

The second type of searches, $B\to K+\rm inv$, corresponds to the scenario when \s is not detected. This may happen if \s decays into invisible particles (such as neutrinos or hypothetical feebly-interacting particles (FIPs) that leave the detector) or if it is too long-lived and escapes the detector before decaying. Therefore, the number of events scales as
\begin{align}\label{eq:invisible-scaling}
     \mathrm{BR}(B^+\to K^+ \varphi)\, \left[\mathrm{BR}(\varphi\to \mathrm{inv}) +  P(r_{\rm det}|\beta\gamma) \,  \mathrm{BR}(\varphi\to f\bar f)\right]\;.
\end{align}
BaBar, Belle, and Belle II already searched for $B\to K+\mathrm{inv}$~\cite{BaBar:2013npw,Belle:2013tnz,Belle:2017oht,Belle-II:2021rof}. 
In the minimal scalar model all scalar decays into, e.g., $e\bar e,\mu\bar\mu,\pi\pi, KK$, etc., would induce a visible activity in the detector. Therefore, the first summand is effectively zero. As for the second contribution, due to the scaling $\tau_{\s}\propto \theta^{-2}$, to be long-lived enough to escape the detector, scalars need to have small $\theta$; otherwise, the probability is exponentially suppressed. The latter means the suppression of the production branching ratio, making ``invisible'' events with scalars very rare. Therefore, the second type of search is not very efficient.

\begin{figure}[bt!]\centering
\includegraphics[width=0.7\textwidth]{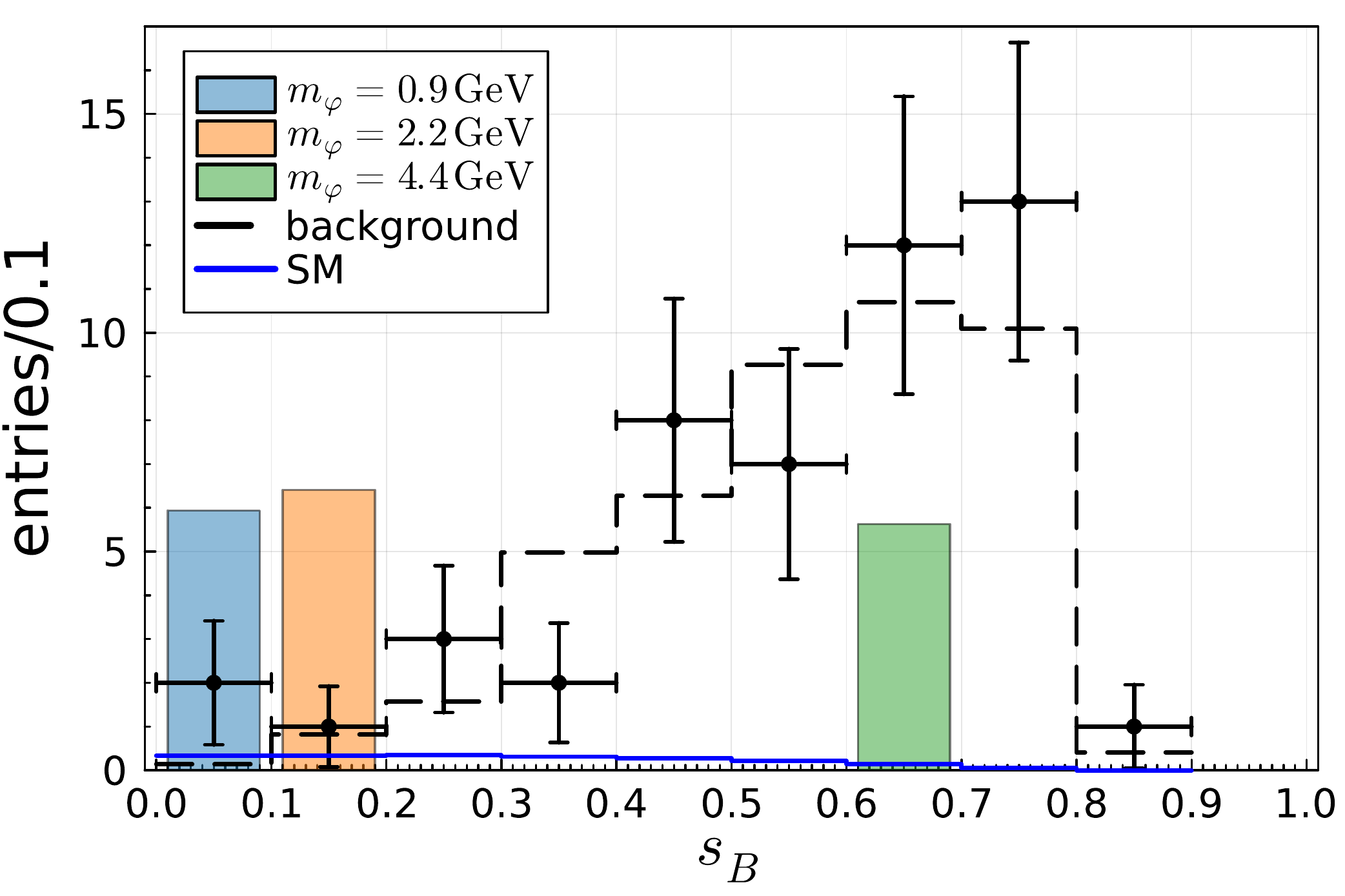}
  \caption{
  The plot shows the experimental BaBar data for $B^+\to K^+ +\mathrm{inv}$~\cite{BaBar:2013npw} binned in the missing invariant mass squared normalised to the decaying $B$ meson mass, $s_B=q^2/m_B^2$. The background is shown as black dashed line and the SM prediction for $B^+ \to K^+\nu\bar\nu$ as a solid blue line. For illustration, we indicate how a light scalar with $\sin\theta=6\times 10^{-3}$ and $\Gamma_{\rm inv}=10$ eV would contribute for different masses $m_\varphi$ using the partial decay width to SM particles calculated following~\cite{Winkler:2018qyg}.}
  \label{fig:BaBar}
\end{figure}

However, the situation may drastically change if the scalar may decay into invisible particles, such that the invisible decay width $\Gamma_{\rm inv}$ becomes comparable with the widths into visible decay states. Similarly to the decays into visible particles, the missing invariant mass distribution would be peaked at $m_{\s}$. This is especially useful since, in addition to the constraint on the branching ratio $\text{Br}(B\to K+\rm inv)$, BaBar~\cite{BaBar:2013npw} provides the distribution in the missing squared invariant mass $q^2$. For 2-body decays that feature a narrow resonance in the $q^2$ distribution, almost all events are contained within one bin. This is illustrated in Fig.~\ref{fig:BaBar}, which shows the number of events in each bin from a scalar with mixing angle $\sin\theta=6\times 10^{-3}$ and invisible decay width $\Gamma_{\rm inv}=10$~eV for different scalar masses $m_\varphi$. This will be used to extract constraints on the scalar mixing angle $\theta$ in Sec.~\ref{sec:results}.

\section{Complementarity of \texorpdfstring{$B\to K \mu\bar\mu$}{B->Kmumu} and \texorpdfstring{$B \to K +\mathrm{inv}$}{B->K+ inv}}
\label{sec:results}

As argued in the previous section, the simple picture changes for scalars with a sizeable invisible decay width. 
Decays to invisible final states leave a signal in $B\to K +\mathrm{inv}$.  
We recast the BaBar search for $B^+ \to K^+ +\mathrm{inv}$~\cite{BaBar:2013npw}, which is the most sensitive search channel of the different $B\to K +\mathrm{inv}$ decays. It provides the data in ten equally spaced bins in the $q^2$ distribution as reproduced in Fig.~\ref{fig:BaBar} from Fig.~5 in~\cite{BaBar:2013npw}. The BaBar data is statistics limited with systematic errors at the percent level. We validated our analysis by reproducing the rescaled SM prediction and the upper bound for BR($B^+\to K^+ \nu\bar\nu$) of the BaBar analysis~\cite{BaBar:2013npw} within 20\% (19\% including systematic errors).\footnote{Our upper limit is lower compared to the one reported in~\cite{BaBar:2013npw}. Improving the analysis would require to generate events to include other cuts such as the one on the $E_{\rm extra}$ variable which quantifies additional energy deposits. Moreover, the precision is also limited by the achievable precision from manually reading off the data from figures 5 and 6 in \cite{BaBar:2013npw}.} As the systematic errors are negligible compared to the statistical errors and the precision of the final result, we neglect them in the following statistical analysis. Following \cite{BaBar:2013npw} we assume the events in each bin are distributed following a Poisson distribution $\mathrm{Poisson}(k|\lambda)=\lambda^k e^{-\lambda}/k!$ 
and thus the likelihood is
\begin{align}
  \mathcal{L} = \prod_{i} \mathrm{Poisson}\left(N_i\Big| \epsilon_i s_i N_{B\bar B}+ b_i \right)
  \end{align}
with the signal efficiency $\epsilon_i~\sim\mathcal{O}(10^{-3})$, extracted from Fig.~6 in~\cite{BaBar:2013npw}, the total number of $B\bar B$ events $N_{B\bar B}=4.71\times 10^8$, and the expected background events in each bin, $b_i$.
The estimates for the background events are separated into peaking background events with a correctly reconstructed tagged event, estimated from Monte Carlo simulations, and combinatorial background from continuum events and incorrectly reconstructed events, which has been extrapolated from data~\cite{BaBar:2013npw}.
Finally, the branching ratio for signal events in each bin is given by 
\begin{equation}
    s_i = \tau_B \int_{\rm bin \,i} \!\!\!\!dq^2 \left(
    \frac{d\Gamma_{\rm SM}(B^+\to K^+\nu\bar\nu)}{dq^2} 
    +\frac{d\Gamma(B^+\to K^+ \varphi(\to\mathrm{inv}))}{dq^2} 
    \right)\;.
\end{equation}
The first term denotes the SM contribution, see~\cite{Altmannshofer:2009ma,Felkl:2021uxi,Becirevic:2023aov}. As the decay width of the scalar is much smaller than its mass for the relevant parameter space, we employ the narrow width approximation\footnote{The detector resolution in $q^2$ is $\mathcal{O}(1\%)$~\cite{BaBar:2013npw}, and thus the broadening due to the finite width $\Gamma_\s$ is negligible.} and find
\begin{align}\label{eq:signal}
\frac{d\Gamma(B^+\to K^+ \varphi(\to\mathrm{inv}))}{dq^2} & = \delta(q^2-m_\varphi^2)\Gamma(B\to K^{+} \varphi)
    \left( \frac{\Gamma_{\rm inv}}{\Gamma_{\varphi,\rm tot}} + P(r_{\rm det}|\beta\gamma) \frac{\Gamma_\varphi^{(\theta)}}{\Gamma_{\varphi,\rm tot}}\right)\;.
\end{align}
The detector size is set to $r_{\rm det}=0.5$m following~\cite{BaBar:2015jvu}. 
For large invisible decay width $\Gamma_{\rm inv}\gg\Gamma_{\s}^{(\theta)}$, the second term is negligible and the differential decay width is proportional to $\sin^2\theta$, while for small invisible decay width the total decay width $\Gamma_{\s,\rm tot} \approx \Gamma_{\s}^{(\theta)} \propto \sin^2\theta$, and thus the $\sin^2\theta$ dependence cancels in the differential decay width. Hence, the constraint from $B^+\to K^++\mathrm{inv}$ can be interpreted as a constraint on $\Gamma_{\rm inv}$ for small invisible decay widths and on $\sin\theta$ for large invisible decay widths.
Without performing any statistical analysis, from the few number of observed events, the total number of $B\bar B$ mesons and the efficiency, we expect to be sensitive to branching ratios BR$(B^+\to K^+\varphi)\sim \mathcal{O}(10^{-5})$.

From the likelihood function, the corresponding $\chi^2$ function is given by 
\begin{align}
  \chi^2 = -2\ln \mathcal{L} = \sum_i f(N_i|\epsilon_i s_i N_{B\bar B}+b_i^{\rm peak} + b_i^{\rm comb})
\end{align}
with $f(n|\nu)=2\nu -2 n\ln \nu  +2 \ln(n!)$. Minimising the $\chi^2$ function with respect to the scalar mixing angle $\sin\theta$ for fixed scalar mass $m_\varphi$ and invisible decay width $\Gamma_{\rm inv}$, we derive an upper limit on $\sin\theta$ at 90\% CL ($\Delta \chi^2=2.7$). 
\begin{figure}[bt!]\centering
\includegraphics[width=0.7\textwidth]{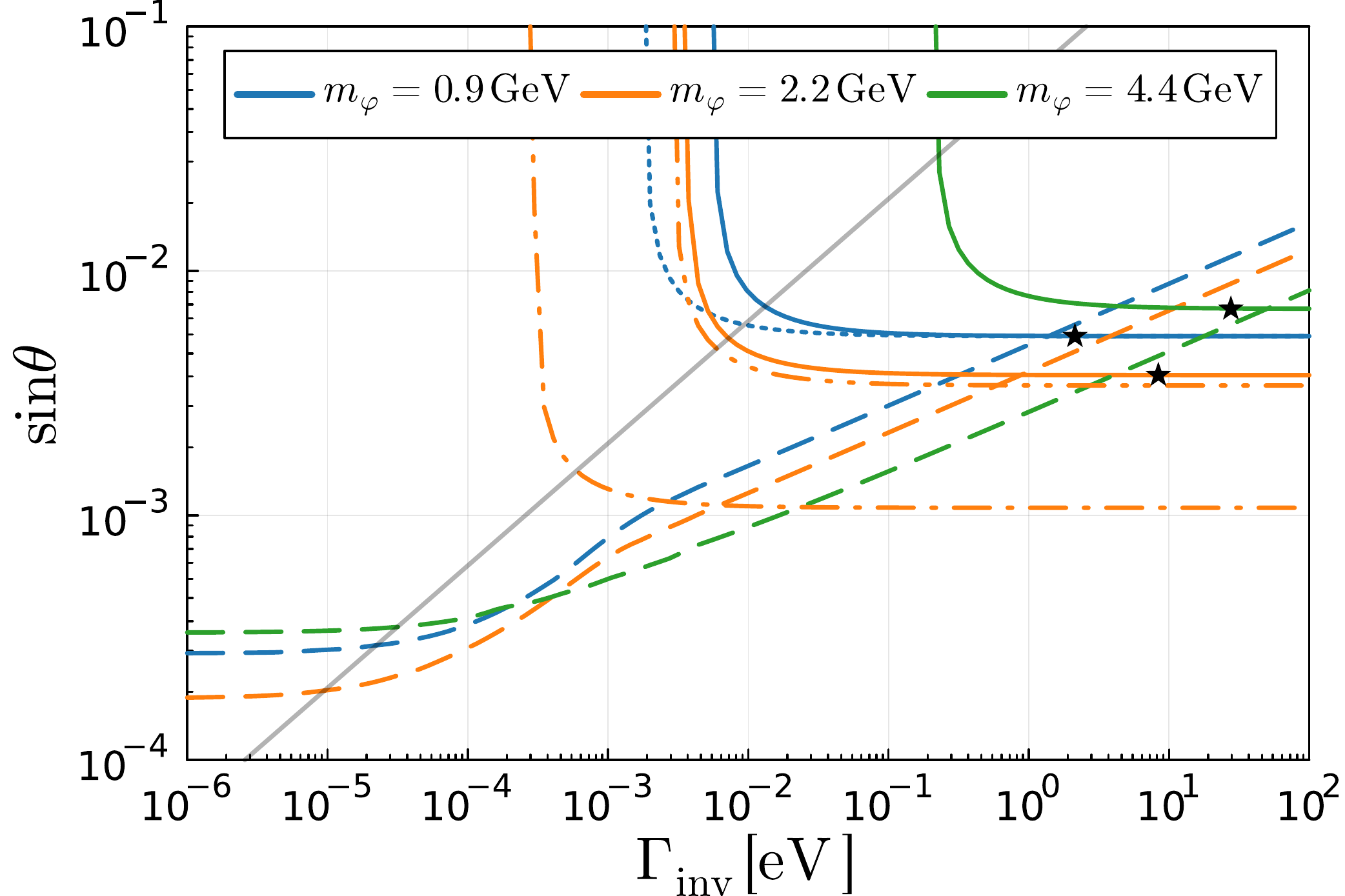}

\caption{
  Exclusion contours for different scalar masses as a function of the invisible decay width $\Gamma_{\rm inv}$. Mixing angles above the dashed lines are excluded by $B^+\to K^+ \varphi (\to \mu\bar\mu)$ from LHCb and above the solid lines by $B^+\to K^+ +\mathrm{inv}$ from BaBar. We use the  prediction for the scalar branching ratios from~\cite{Winkler:2018qyg} with included NLO corrections. The blue dotted line shows the BaBar constraint based on \cite{Gorbunov:2023lga}.
  The double-dot-dashed (dot-dashed) orange line for $m_\varphi=2.2$~GeV corresponds to BR($B^+ \to K^+ \varphi$) being equal to (10\%  of) the SM prediction. The grey line indicates $\Gamma_{\rm inv}=\Gamma_{\s}^{(\theta)}$ for $m_\varphi=2.2$ GeV. The stars indicate the maximum invisible decay width $\Gamma_{\rm inv}=\Gamma(\varphi\to NN)$ in the gauged $B-L$ model. 
  }
  \label{fig:gamma-sintheta}
\end{figure}

\begin{figure}[bt!]\centering
\includegraphics[width=0.5\textwidth]{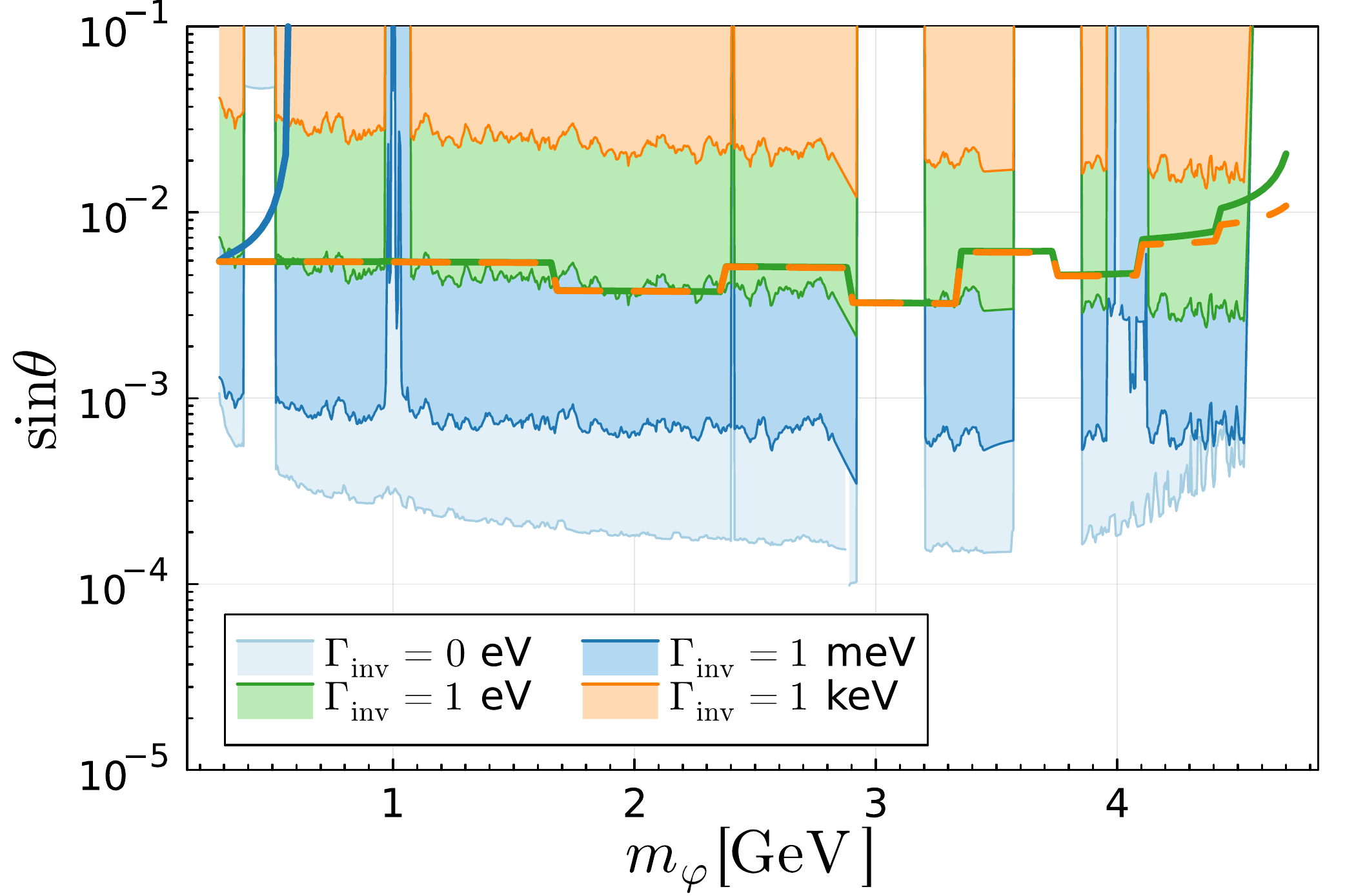}%
\includegraphics[width=0.5\textwidth]{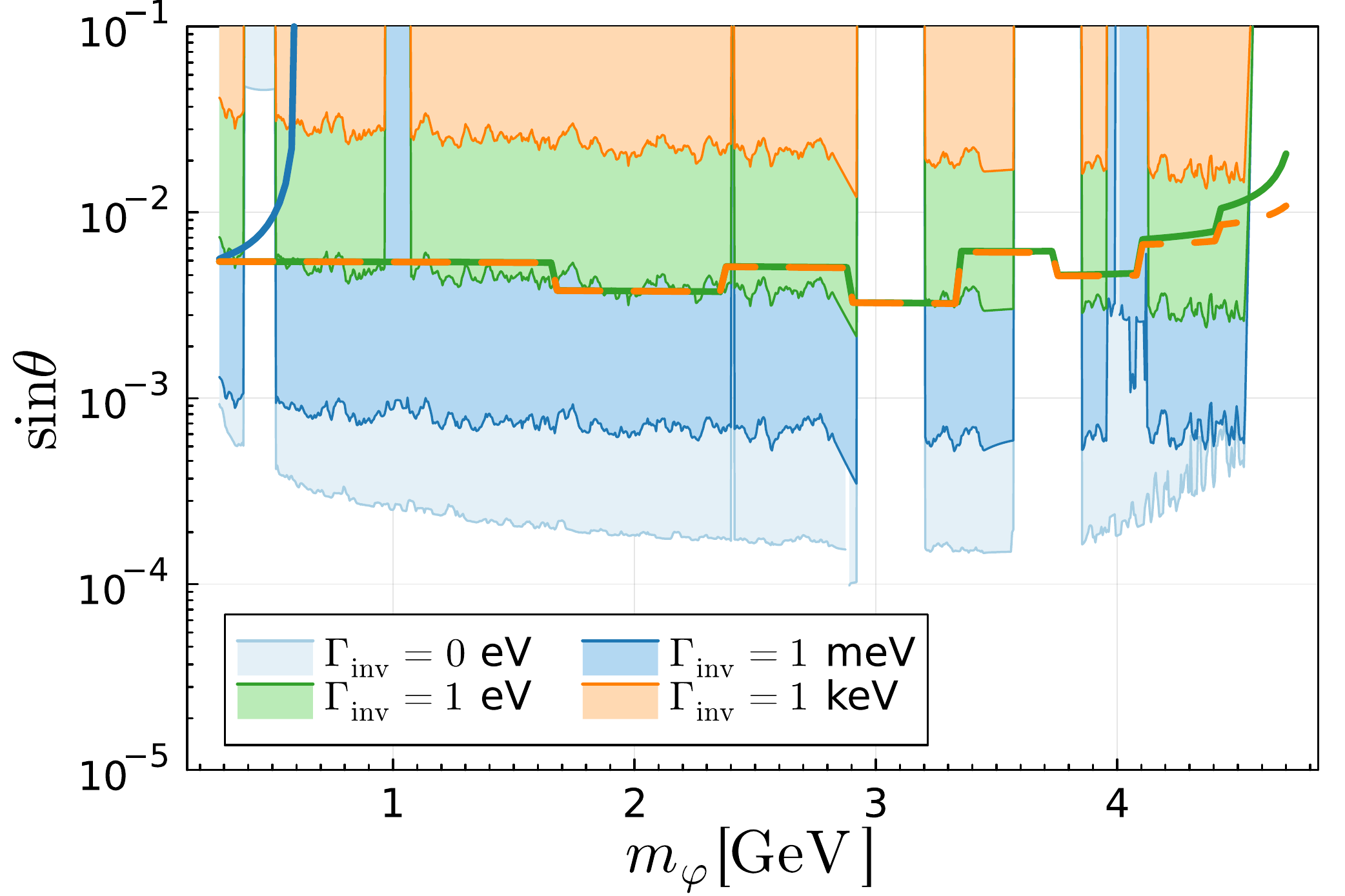}%

\caption{
  Recast LHCb exclusions (shaded regions) for a Higgs-like scalar for selected fixed invisible decay widths in comparison to the upper bounds on the mixing angle $\theta$ from the BaBar search for $B^+\to K^+ +\mathrm{inv}$ for the same invisible decay widths using solid lines with the same colours. The partial decay width to SM particles in the left (right) plot has been calculated following \cite{Winkler:2018qyg} (\cite{Gorbunov:2023lga}). 
  }
  \label{fig:mphi-sintheta}
\end{figure}

The main results are presented in Figs.~\ref{fig:gamma-sintheta} and~\ref{fig:mphi-sintheta}, which show the constraints on the scalar mixing angle $\sin\theta$ from the search for $B^+\to K^+ \varphi(\to \mu\bar\mu)$ at LHCb~\cite{LHCb:2016awg} and invisible decay searches at BaBar~\cite{BaBar:2013npw} as a function of the invisible decay width $\Gamma_{\rm inv}$ and the scalar mass $m_\varphi$ for different benchmark values. The BaBar constraints are shown as solid contours in both figures, while the LHCb constraints are indicated by dashed lines in Fig.~\ref{fig:gamma-sintheta} and by shaded regions in Fig.~\ref{fig:mphi-sintheta}. 
For small invisible decay widths $B^+\to K^+ \varphi (\to \mu\bar\mu)$ places the most stringent constraint on $\sin\theta$, while $B^+\to K^+ \varphi(\to \mathrm{inv})$ places the most stringent constraint for large invisible decay widths. The  cross over occurs for $\Gamma_{\rm inv} \sim \mathcal{O}(1-50)$ eV depending on the scalar mass. 
 
The dependence of the LHCb search for $B^+\to K^+\varphi(\to\mu\bar\mu)$~\cite{LHCb:2016awg} is straightforward to understand. For $\Gamma_{\rm inv}\ll \Gamma_{\s}^{(\theta)}$, there is no dependence on the invisible decay width, as can be observed on the left side of Fig.~\ref{fig:gamma-sintheta}. The grey line shows $\Gamma_{\rm inv} =\Gamma_\s ^{(\theta)}$ for $m_\varphi=2.2$ GeV (orange contours). At the intersection of the dashed orange and grey line, the LHCb constraint starts to weaken. For $\Gamma_{\rm inv} \gtrsim 1$ meV, the LHCb constraints are described by lines on a log-log scale in Fig.~\ref{fig:gamma-sintheta}, because the scalar decays promptly, while for $\Gamma_{\rm inv}\lesssim 1$ meV, the constraint depends on the finite scalar lifetime. Similarly in Fig.~\ref{fig:mphi-sintheta}, the constraint for $\Gamma_{\rm inv}=1$ keV can be obtained from the one for $\Gamma_{\rm inv}=1$ eV by rescaling, while the constraints for $\Gamma_{\rm inv}\lesssim 1$ meV feature a non-trivial dependence on the invisible decay width $\Gamma_{\rm inv}$ via the dependence on the scalar lifetime $\tau_{\s}$. Note, the LHCb result is largely insensitive to the different predictions~\cite{Winkler:2018qyg,Gorbunov:2023lga} for the branching ratios.

The dependence of the BaBar constraints (solid contours) in Fig.~\ref{fig:gamma-sintheta} can be understood from the above argument: For large $\Gamma_{\rm inv}\gg \Gamma_{\s}^{(\theta)}$, $B\to K+\mathrm{inv}$ can be interpreted as a constraint on $\sin\theta$, while for small $\Gamma_{\rm inv}$ it has to be interpreted as a constraint on $\Gamma_{\rm inv}$, which explains the sharp drop in sensitivity for $\Gamma_{\rm inv}\lesssim 1$ meV. This can also be observed in Fig.~\ref{fig:mphi-sintheta}: The BaBar constraints for $\Gamma_{\rm inv}=1$ eV and $1$ keV agree except for scalar masses close to the kinematic cutoff. There is no sensitivity for $\Gamma_{\rm inv}=0$ eV and  we observe a strong dependence on the scalar mass for $\Gamma_{\rm inv}=1$ meV, because the BaBar sensitivity depends on the $q^2$ bin.
To illustrate the sensitivity of $B^+\to K^+ +\mathrm{inv}$ to new physics, we show iso-contours
for the branching ratio of BR($B^+\to K^+\varphi)$ being equal to (10\% of) the SM branching ratio BR($B^+\to K^+ \nu\bar\nu$) for a scalar with $m_{\s}=2.2$ GeV as dot-dashed (dot-dot-dashed) orange lines in Fig.~\ref{fig:gamma-sintheta}.

Finally, depending on the scalar mass,  
the theoretical uncertainty in the scalar's hadronic decay width affects the lower bounds for visible/invisible signatures.
Let us first consider the invisible case. In Fig.~\ref{fig:gamma-sintheta}, we consider two descriptions of the scalar's width discussed in Sec.~\ref{sec:model} for the mass $m_{\s} = 0.9$ GeV: as in~\cite{Winkler:2018qyg} (the solid blue line) and~\cite{Gorbunov:2023lga} (the dotted blue line). Also, Fig.~\ref{fig:mphi-sintheta} shows the sensitivity from the invisible signature assuming these two descriptions (the left and right panels correspondingly). In the domain of large $\Gamma_{\rm inv}\gg \Gamma^{(\theta)}_{\s}$, where $\Gamma^{(\theta)}_{\s}$ is the total width controlled by $\theta$, the dependence of the number of events on the total width disappears. Therefore, the sensitivity does not depend on the uncertainty, as we see for $\Gamma_{\rm inv}$. Indeed, in Eq.~\eqref{eq:invisible-scaling}, the first summand in the brackets reduces to 1, while the second summand tends to 0. 
Once $\Gamma_{\rm inv}$ decreases, $\Gamma^{(\theta)}_{\s}$ becomes essential. In particular, in the limit $\Gamma_{\rm inv}\ll \Gamma^{(\theta)}_{\s}$, the number of events scales as $(\Gamma^{(\theta)}_{\s})^{-1}$: larger width means lower number of events. The width from~\cite{Winkler:2018qyg} is resonantly enhanced compared to the one from~\cite{Gorbunov:2023lga}, and therefore the sensitivity of the invisible signature is weaker in the former case.

For the visible case, the number of events scales as given in Eq.~\eqref{eq:visible-scaling}. We illustrate the results for two different descriptions in Fig.~\ref{fig:mphi-sintheta}. As we already discussed, if the invisible width is very small or zero, the LHCb bounds extend to small values of the mixing angle where the probability of the scalar to decay inside the decay cancels with the $\text{Br}(\s\to \mu\bar\mu)$. In the opposite case of large $\Gamma_{\rm inv}$, the decay probability does not have such scaling and tends to 1. The dependence on the total width via $\text{Br}(\s\to \mu\bar\mu)$ survives. 
Similarly to the invisible signature, larger $\Gamma^{(\theta)}_{\s}$ means weaker bounds. In particular, assuming the description from~\cite{Winkler:2018qyg} and $\Gamma_{\rm inv} = 1\text{ eV}$, we see that the sensitivity is reduced at $m_\varphi=1$ GeV due to the resonant enhancement of the $f_{0}(980)$ and thus large visible decay width into SM particles via the scalar mixing. The shape is different if assuming the description as in~\cite{Gorbunov:2023lga}, where no resonant enhancement exists.

\section{SM extensions with a light scalar with invisible width}
\label{sec:examples}

There are several possibilities how the light scalar may decay and escape detection in the LHCb searches for $B\to K \varphi (\to\mu\bar\mu)$. An attractive possibility is to couple the scalar field to fermionic dark matter. However thermal production of this dark matter candidate in the early universe is strongly constrained~\cite{Krnjaic:2015mbs} and production via freeze-in requires small couplings. We thus focus on scalar decays to unstable particles which further decay to lighter SM particles, like HNLs. 
HNLs are well-motivated as explanation of neutrino masses via the seesaw mechanism~\cite{Minkowski:1977sc}.
Big bang nucleosynthesis (BBN) places a lower bound on the lifetime of GeV-scale HNLs of $\tau_N<0.02$s~\cite{Boyarsky:2020dzc} and thus a lower bound on the active-sterile mixing angle. Direct searches on the other hand provide an upper bound on the active-sterile mixing angle. Assuming that the HNLs generate neutrino masses, together the two constraints exclude HNL masses below $0.33-0.36$ GeV, set by the kinematic threshold of $K\to \pi \nu N$, apart from a small window $M_N\in [0.12,0.14]$~GeV~\cite{Bondarenko:2021cpc}.
For the allowed parameter space GeV-scale HNLs escape the detector undetected, see e.g.~\cite{Bondarenko:2021cpc,Bolton:2022tds}. 

\subsection{Real scalar coupling to heavy neutral leptons}

The simplest viable scenario is to couple a real scalar field to HNLs $N_i$, which allows to generate the invisible scalar decay width via $\varphi\to NN$. The relevant Lagrangian is
\begin{align}
  \mathcal{L} & = - \frac12  N^TC (\mu_N + y_N \varphi)  N + \mathrm{h.c.} 
\end{align}
where $C$ is the charge conjugation matrix, $M_N = \mu_N+ y_N v_\phi$ is the HNL mass term and $y_N$ the Yukawa coupling to the scalar $\varphi$. Hence, the Yukawa coupling $y_N$ and the HNL mass can be chosen independently. 

An important constraint on this scenario comes from SM Higgs decays. In presence of a quartic Higgs portal interaction  $\tfrac{\lambda_{H\varphi}}{2} H^\dagger H \varphi^2$, the SM Higgs can decay in the real scalar with the branching ratio
\begin{equation}
  \mathrm{BR}(h\to \varphi\varphi) = \frac{\lambda^{1/2}(m_h^2,m_\varphi^2,m_\varphi^2)}{32 \pi \sqrt{2}G_F m_h^3 \Gamma_h} |\lambda_{H\varphi}|^2 
\end{equation}
with the K\"all\'en function $\lambda(x,y,z)=x^2+y^2+z^2-2xy-2xz-2yz$.
The upper bound BR($h\to \mathrm{inv})\leq 0.18$~\cite{CMS:2022qva} constrains the quartic coupling as $|\lambda_{H\varphi}|\lesssim 0.01$. Similarly, for non-zero vacuum expectation value (VEV) of $\varphi$, the quartic Higgs portal interaction contributes to the scalar mixing. Demanding this contribution to be small results in $\lambda_{H\varphi}\ll \sqrt{2} \sin\theta m_h^2/vv_\varphi$ with the electroweak VEV $v^2 = 1/\sqrt{2} G_F$, i.e.~we assume that the mixing is dominated by the cubic term in the Higgs potential. 

Even in absence of a quartic Higgs portal interaction, there is a contribution to invisible Higgs decay from $h\to NN$  which translates into an upper bound on the HNL Yukawa coupling
\begin{align}
  \mathrm{BR}(h\to N N) = \sum_{i,j}\frac{|y_{Nij}|^2 \sin^2 \theta}{8\pi(1+\delta_{ij})} \frac{m_h}{\Gamma_h} 
  \qquad \Rightarrow\qquad
  \sum_{i,j} \frac{|y_{Nij}|^2}{1+\delta_{ij}} \lesssim 
  1.4 \,\left(\frac{10^{-2}}{\sin\theta}\right)^2\;. 
\end{align}

The partial decay width for scalar decay to HNLs is 
\begin{align}\label{eq:GammaphiNN}
  \Gamma(\varphi\to N_iN_j)&= \frac{m_\varphi   |y_{Nij}|^2 }{8\pi(1+\delta_{ij})}   
 \lambda^{1/2}(1,x_i^2,x_j^2)
 (1-(x_i+x_j)^2) 
\end{align}
with $x_i\equiv M_{Ni}/m_\varphi$. 
Taking the constraint from invisible Higgs decay into account, we obtain an upper bound for the invisible width for $x_i\ll 1$
\begin{align}
  \Gamma(\varphi\to NN)&\lesssim 0.06  \,\left(\frac{10^{-2}}{\sin\theta}\right)^2\, m_\varphi \;.
\end{align}
Hence, there is no strong constraint on the partial decay width $\Gamma(\varphi\to NN)$ and a sizeable invisible decay width is allowed for a real scalar coupling to HNLs. As argued above, the lifetime of HNLs is long compared to the size of the detector and thus they escape undetected. For example, using the lifetime of HNLs from~\cite{Bondarenko:2018ptm}, we explicitly find for $m_N=2.3$ GeV a decay length of \mbox{$c\tau_N=3.75\times 10^5$ m} at the seesaw line $U^{2}_{\text{seesaw}} \sim 5\cdot 10^{-11} (1\text{ GeV}/m_{N})$~\cite{Alekhin:2015byh}. Together with the allowed mass range of HNLs, this demonstrates the whole range of invisible partial decay widths in Fig.~\ref{fig:gamma-sintheta} can be obtained in the real singlet model. The model is also able to explain light neutrino masses via the seesaw mechanism~\cite{Minkowski:1977sc}. 

\subsection{\texorpdfstring{$B-L$}{B-L} gauge symmetry}

The real scalar field discussed in the previous subsection has many undetermined parameters, which can be reduced by introducing a symmetry. A well-motivated scenario is the $B-L$ symmetry. After spontaneous breaking of the $B-L$ symmetry, a Majorana mass term for the HNLs is generated, which provides an explanation of active neutrino masses via the seesaw mechanism~\cite{Minkowski:1977sc}. 
A global $B-L$ symmetry is straightforwardly ruled out: spontaneous breaking of the $B-L$ symmetry results in a Majoron, which is efficiently produced via its interactions with the HNLs. It thus contributes to the effective number of neutrinos, $N_{\rm eff}$, and as it only decouples below the scale of the HNLs, its contribution to $N_{\rm eff}$ is too large and excluded by BBN. These constraints are avoided by promoting the global $B-L$ symmetry to a gauge symmetry, which has been first proposed in~\cite{Davidson:1978pm,Marshak:1979fm}. 

The relevant interactions for the following discussion are  
\begin{equation}
  \mathcal{L} \supset (D_\mu\phi)^\dagger ( D^\mu \phi)
  - \lambda_{H\phi}\left(|\phi|^2-\frac{v_\phi^2}{2}\right) \left(H^\dagger H -\frac{v^2}{2}\right)
  -\frac12 N^TC y_N \phi N 
  \;.
\end{equation}
After spontaneous breaking of the $B-L$ gauge symmetry with $\langle \phi \rangle = v_\phi/\sqrt{2}$, HNLs and the $Z'$ gauge boson acquire the masses\footnote{Without loss of generality, we use mass eigenstates for the HNLs. The Yukawa coupling matrix $y_N$ is then approximately diagonal, where off-diagonal entries only appear due to small non-zero active-sterile neutrino mixing, which we neglect in the following discussion.} $M_{Ni}= y_{Ni} v_\phi/\sqrt{2}$ and $m_{Z'}= 2 g_{\rm BL} v_\phi$ with the $B-L$ gauge coupling. The interactions of the real scalar $\varphi \equiv \mathrm{Re}(\phi)/\sqrt{2}$ are by construction proportional to the HNL and $Z'$ masses 
\begin{equation}\label{eq:phiZZ}
  \mathcal{L} \supset - \frac12 \varphi N_i^T C\frac{M_{Ni}}{v_\phi} N_i +  \frac12 \frac{m_{Z'}^2}{v_\phi^2}\varphi^2 Z^\prime_\mu Z^{\prime\mu} + \frac{m_{Z'}^2}{v_\phi}\varphi Z^\prime_\mu Z^{\prime\mu}
\end{equation}
which results in tight connections between the different observables.

The partial decay width for decay to HNLs in Eq.~\eqref{eq:GammaphiNN} can be expressed in terms of the masses and the gauge coupling $g_{\rm BL}$
\begin{align}
  \Gamma(\varphi\to NN)
  = \frac{g_{\rm BL}^2 m_\varphi^3  }{2\pi m_{Z'}^2} \sum_i x_i^2   (1-4x_i^2)^{3/2} 
\end{align}
and the decay to a pair of $Z'$ gauge bosons is given in terms of 
\begin{align}
  \Gamma(\varphi\to Z'Z') 
 = \frac{g_{\rm BL}^2 m_\varphi}{8\pi} \frac{(1-4z)^{1/2}\left(1 -4z +12z^2 \right)}{z}
\end{align}
with $z=m_{Z'}^2/m_\varphi^2$. The $Z'$ gauge boson however is not stable and further decays to SM particles. Its lifetime is inversely proportional to the square of the gauge coupling, $\tau_{Z'} \propto g_{\rm BL}^{-2} m_{Z'}^{-1}$. Hence, the $Z'$ gauge boson quickly decays to neutrinos, and depending on its mass to charged leptons and hadrons. The standard model charged leptons couple to $Z'$ vectorially and the branching ratio for active neutrinos can be straightforwardly obtained as
\begin{align}
  \mathrm{Br}(Z' \to \ell \bar\ell)
  =\frac{g_{\rm BL}^2}{12\pi}\frac{m_{Z'}}{\Gamma_{Z'}}\left(1-4y\right)^{1/2}\left(1+2 y\right)
  \quad\mathrm{and}\quad
  \mathrm{Br}(Z' \to \sum_i\nu_i \bar\nu_i)
  =\frac{g_{\rm BL}^2}{8\pi}\frac{m_{Z'}}{\Gamma_{Z'}}
\end{align}
with $y=m_\ell^2/m_{Z'}^2$. Hence, a substantial fraction of the $Z'$ gauge bosons decays to visible final states which leave a signal in the detector with multiple leptons and/or hadrons and thus do not contribute to either $B^+\to K^+\varphi(\to \mu\bar\mu)$ or $B^+\to K^+ \varphi(\to \mathrm{inv})$.
Variants of the $B-L$ gauge symmetry with non-universal lepton number, see e.g.~\cite{Kownacki:2016pmx}, may forbid $Z'$ decays to light leptons and thus evade any constraints from the cascade decay $B^+\to K^+ \varphi (\to Z' Z'\to \ell \bar\ell \ell\bar\ell)$, see e.g.~\cite{CidVidal:2022mrx} for a discussion of a decay with two muon-antimuon pairs in the final state. Final states with multiple muons have also been studied in \cite{Blance:2019ixw}. In the following we consider the scenario, where the decay $\varphi\to Z'Z'$ is kinematically forbidden by choosing $m_{Z'}\geq m_\varphi/2$.

\begin{figure}[tb!]\centering
\includegraphics[width=0.7\textwidth]{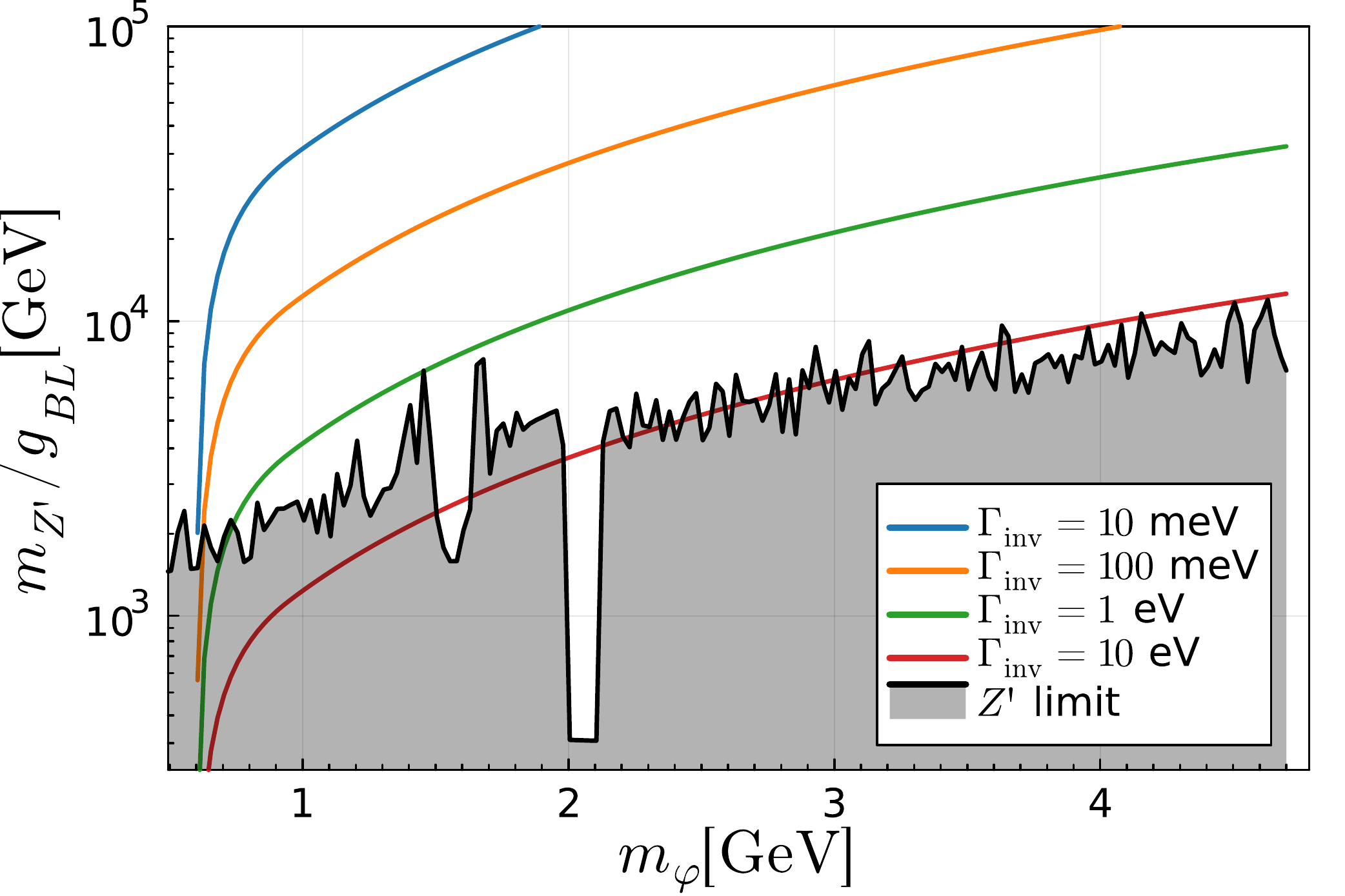}
\caption{Maximum decay width $\Gamma(\varphi\to NN)$ as function of $m_\varphi$ and $m_{Z'}/g_{\rm BL}$. The grey-shaded region is excluded by $Z'$ searches assuming $m_{Z'}\geq m_\varphi/2$  such that $\varphi\to Z'Z'$ is kinematically forbidden. 
}
\label{fig:BLgamma}
\end{figure}

In Fig.~\ref{fig:BLgamma} we present the maximum decay width $\Gamma(\varphi\to NN)$ as a function of the scalar mass $m_\varphi$ and the $Z'$ mass $m_{Z'}/g_{\rm BL}$. For this we consider two degenerate HNLs with mass $m_N=\max(m_\phi/\sqrt{10},0.3\, \mathrm{GeV})$ which maximises the partial decay width $\Gamma(\varphi\to NN)$ and respects the lower bound on the mass of HNLs.
The grey-shaded region is excluded by $Z'$ searches.
It has been extracted from the top-left plot of Fig.~2 in \cite{Kling:2020iar} (see also \cite{Ilten:2018crw,Bauer:2018onh}). The coloured solid contours show $\Gamma_{\rm inv} \in [0.01, 0.1,1 ,10]$ eV.  Together this demonstrates the possibility of a sizeable invisible scalar decay width $\Gamma_{\rm inv}\gtrsim 1$ eV. We indicate the maximum possible invisible decay width consistent with the constraint on $m_{\rm Z'}/g_{\rm BL}$ also in Fig.~\ref{fig:gamma-sintheta} as stars for the three benchmark masses.

Finally, invisible Higgs decays place a constraint on the gauge $B-L$ model via the quartic Higgs portal $H^\dagger H \phi^\dagger\phi$.
For small mixing and light scalars, $m_\varphi\ll m_h$, the Higgs branching ratio to a pair of scalars is given by  
\begin{equation}
 \mathrm{BR}(h\to \varphi\varphi)
  = \frac{ g_{\rm BL}^2}{8\pi} \frac{ m_h^3 } {m_{Z'}^2 \Gamma_h}  \sin^2\theta \;,
\end{equation}
where we expressed the Higgs portal coupling in terms of the scalar mixing angle $\theta$.
Higgs decays to HNLs and $Z'$ gauge bosons are negligible in this case, because they are suppressed by the small masses of the final state particles. The constraint on the invisible Higgs decay branching ratio thus results in an upper bound on $\sin\theta$,
\begin{equation}
    \sin^2\theta \leq 0.93 \left(\frac{m_{Z'}}{\mathrm{GeV}} \right)^2 \left(\frac{10^{-4}}{g_{\rm BL}}\right)^2\;.
\end{equation}
It is however weaker than the searches for the scalar in $B$ meson decays and does not pose any additional constraint.
%


\section{Conclusions}
\label{sec:conclusions}
We considered a scalar, which mixes with the SM Higgs boson with an additional invisible decay width $\Gamma_{\rm inv}$, and scrutinised the constraints on the scalar mixing angle from the search $B^+\to K^+ \varphi (\to\mu \bar\mu)$ at LHCb~\cite{LHCb:2016awg}. The LHCb search looses sensitivity for invisible decay widths larger than the visible decay width,  $\Gamma_{\rm inv} \gtrsim \Gamma_{\s}^{(\theta)}$. We demonstrated that for scalars with an invisible decay width of $\Gamma_{\rm inv}\gtrsim\mathcal{O}(1-50)$ eV, the decay $B^+\to K^+ \varphi(\to\mathrm{inv})$ provides the most stringent constraint on the scalar mixing angle $\theta$, which opens an opportunity for Belle II to discover new physics in $B^+\to K^+ +\mathrm{inv}$.

We provided two explicit models which realise a sizeable invisible decay width. The gauged $B-L$ model is well motivated as an explanation of non-zero neutrino masses. In this model, the scalar breaking $B-L$ gauge symmetry may decay to heavy neutral leptons.
The heavy neutral leptons are long-lived on the time scale of the detector, escape it undetected and thus contribute to the invisible decay width of the $B-L$ scalar. 
This scenario is mostly constrained by searches for the $Z'$ gauge boson which limits the invisible decay width to less than $\mathcal{O}(30)$ eV. As those constraints do not apply in the real scalar model, it is possible to obtain an invisible decay width in the MeV range. 

Finally, as the phenomenology mainly depends on scalar mixing and the coupling of the scalar to heavy neutral leptons, the conclusions from the $B-L$ model apply at least qualitatively to other gauged $U(1)'$ extensions of the SM,
as long as the scalar spontaneously breaking the $U(1)'$ symmetry can decay to heavy neutral leptons.

\section*{Acknowledgements}
MS thanks the Karlsruhe Institute of Technology for their hospitality and acknowledges support by the Australian Research Council Discovery Project DP200101470. 
This work was supported by the KIT International Excellence Fellowships Program with funds granted to the University of Excellence concept of the Karlsruhe Institute of Technology.  
This work has been supported by the European Union’s Framework Programme for Research and Innovation Horizon 2020 under grant H2020-MSCA-ITN-2019/860881-HIDDeN.

\appendix
\section{Relevant decay channels \texorpdfstring{$B\to K^{(*)} \varphi$}{B->K(*)φ}}
\label{app:formfactor}

At 1 loop, the SM Higgs $h$ has flavour-violating couplings to quarks~\cite{Willey:1982mc,Leutwyler:1989xj}\footnote{See \cite{Kachanovich:2020yhi} for a calculation of the effective vertex of the second Higgs in $R_\xi$ gauge.} from electroweak corrections. The dominant contribution originates from top quarks 
\begin{align}
  \mathcal{L}_{\rm eff} =  \frac{h}{v} C_{sb} m_b \overline{s} P_R b + \mathrm{h.c.} 
  \qquad\mathrm{with} \quad C_{sb} = \frac{3\sqrt{2} G_F m_t^2 V_{ts}^* V_{tb}}{16\pi^2}\;.
\end{align}
They give rise to various decays $B\to X_s + \s$, where $X_{s}$ are mesons containing a strange ($s$) quark. The matrix element of the process separates into short-distance part $C_{sb} m_b\sin\theta /v$ and the matrix element $\mathcal{M}_{BX_s}$ which describes the long-distance QCD contributions
\begin{equation}
    \mathcal{M}_{B\to \s +X_{i}} = \frac{C_{sb}m_b \sin\theta}{v} \mathcal{M}_{BX_{s}}, \quad \mathcal{M}_{BX_{s}} = \langle X_{s}|\bar{s}P_{R}b| B\rangle
    \;.
\end{equation}
The matrix elements $\mathcal{M}_{BX_{s}}$ can be expressed in terms of the matrix elements mediating the weak charged current transitions
\begin{equation}
    \mathcal{M}_{BX_{s},V} = \langle X_{s}|\bar{s}\gamma_{\mu}b|B\rangle, \quad  \mathcal{M}_{BX_{s},A} = \langle X_{s}|\bar{s}\gamma_{\mu}\gamma_{5}b|B\rangle \;,
\end{equation}
see Appendix F in~\cite{Boiarska:2019jym} for a detailed discussion how to express the matrix elements for different states $X_s$ (scalar, pseudoscalar, vector, axial-vector, or tensor) in terms of form-factors.

We focus on $B$ decays into $K^{+}$, $K_S^0$ and $K^*$ mesons.
The decay width for $B^+\to K^+\varphi$ is\footnote{The production of a light Higgs from flavoured meson decays has been first considered in~\cite{Willey:1982mc,Leutwyler:1989xj}.}
\begin{align}\label{eq:BKphi}
\Gamma(B^+\to K^+ \varphi )  & =
\frac{|C_{sb}|^2 \sin^2\theta \sqrt{2} }{64\pi} \frac{G_F (m_B^2-m_K^2)^2 }{m_B} \frac{m_b^2}{(m_b-m_s)^2} |f_0^{BK}(m_\varphi^2)|^2 \lambda^{1/2}\left(1,\frac{m_K^2}{m_B^2},\frac{m_\varphi^2}{m_B^2}\right)
\end{align}
with the K\"all\'en function $\lambda(x,y,z)=x^2+y^2+z^2-2xy-2xz-2yz$, and $f_0^{\rm BK}$ is the transition form-factor $B\to K$.  

The decay width for the neutral $B$ meson to $K_S^0$ is obtained from Eq.~\eqref{eq:BKphi} by replacing the charged meson masses with neutral meson masses and dividing the partial decay width by two to take into account the overlap of $K_S^0$ and $K^0$. The partial decay widths for $K^*$ vector mesons in the final state are given by
\begin{align}
\label{eq:BKstarphi}
\Gamma(B\to K^* \varphi )  & = \frac{|C_{sb}|^2 \sin^2\theta \sqrt{2}}{64\pi}  G_F m_B^3 \frac{m_b^2}{(m_b+m_s)^2 } \left|A_0^{BK^*}(m_\varphi^2)\right|^2\lambda^{3/2}\left(1,\frac{m_{K^*}^2}{m_B^2},\frac{m_\varphi^2} {m_B^2}\right)\;,
\end{align}
where $K^{*}$ refers either to the charged $K^{*+}$ or neutral $K^{*0}$. The $B\to K^{(*)}$ form factors are parameterised by (see e.g.~\cite{Gubernari:2018wyi,Gubernari:2023puw})\footnote{See~\cite{Gubernari:2020eft,Gubernari:2022hxn} for technical details about non-local form factors for $B\to K^{(*)}\mu\bar\mu$.}
\begin{align}
F(q^2) &= \frac{\sum_{i=0}^2 a_i ( z(q^2)-z(0))^i}{1-q^2/m_R^2},
&
z(t) & = \frac{\sqrt{t_+-t}-\sqrt{t_+-t_0}}{\sqrt{t_+-t}+\sqrt{t_+-t_0}}
\end{align}
with $t_\pm = (m_B \pm m_K)^2$, $t_0 = t_+ (1-\sqrt{1-t_-/t_+})$. The resonance mass $m_R$ and the coefficients $a_i$ for the two form factors $f_0^{BK}$ and $A_0^{BK^*}$ are collected in Tab.~\ref{tab:BKformfactor}. Due to finite width of the $K^*$, the partial decay width $\Gamma(B\to K^*(\to K\pi)+\mathrm{inv})$ is enhanced by 20\%~\cite{Descotes-Genon:2019bud}. 
\begin{table}[t!]
    \centering
    \begin{tabular}{lcccc}
    \toprule
process &    $m_R$ [GeV] & $a_0$ & $a_1$ & $a_2$ \\\midrule
  $B \to K$ &   $5.630$  &$0.33196382$ & $0.33478906$ & $0.00372445$\\
  $B \to K^*$ &   $5.336$  & $0.34213768$ & $-1.14740932$  & $2.37275841$\\\bottomrule
    \end{tabular}
  \caption{Parameters for the $B\to K^{(*)}$ scalar form factors~\cite{Gubernari:2023puw} $f_0^{BK}$ and $A_0^{BK^*}$ in the BSZ parametrization~\cite{Bharucha:2015bzk}.}
    \label{tab:BKformfactor}
\end{table}

\setlength{\bibsep}{.2\baselineskip plus 0.3ex}
\bibliographystyle{utphys} 
\bibliography{refs}
\end{document}